\def\year{2022}\relax
\documentclass[letterpaper]{article} % DO NOT CHANGE THIS
\usepackage{aaai22}  % DO NOT CHANGE THIS
\usepackage{times}  % DO NOT CHANGE THIS
\usepackage{helvet}  % DO NOT CHANGE THIS
\usepackage{courier}  % DO NOT CHANGE THIS
\usepackage[hyphens]{url}  % DO NOT CHANGE THIS
\usepackage{graphicx} % DO NOT CHANGE THIS
\usepackage{natbib}  % DO NOT CHANGE THIS AND DO NOT ADD ANY OPTIONS TO IT
\usepackage{caption} % DO NOT CHANGE THIS AND DO NOT ADD ANY OPTIONS TO IT
\DeclareCaptionStyle{ruled}{labelfont=normalfont,labelsep=colon,strut=off} % DO NOT CHANGE THIS
\usepackage{amsmath}
\usepackage{amsthm}
\newtheorem{lemma}{Lemma}
\newtheorem{theorem}{Theorem}
\newtheorem{example}{Example}

\newtheorem{definition}{Definition}
\newtheorem{corollary}{Corollary}
\usepackage{amssymb}
\usepackage{todonotes}
\usepackage{breqn}
\usepackage{colortbl}
\usepackage{multirow}
\usepackage{arydshln}
\usetikzlibrary{arrows}
\usetikzlibrary{shapes}
\usepackage{bm}
\usepackage{versions}
\usepackage[bbgreekl]{mathbbol}
\usepackage{textgreek}
\usepackage{thmtools,thm-restate}

\def\mc{\mbox{\sc mc}}

\def\L{\mathcal{L}}
\def\bx{\bm{x}}

\def\fomc{\mbox{\sc fomc}}

\def\wfomc{\mbox{\sc wfomc}}

\def\bh{\bm{h}}
\def\kh{\bk,\bh}

\def\CC{\mathbb{C}}
\def\bk{{\bm{k}}}

\def\cgreen{\cellcolor{green!10}}
\def\cred{\cellcolor{red!10}}
\def\card#1{{\left|#1\right|}}

\title{Weighted Model Counting in FO$^2$ with Cardinality Constraints and 
Counting Quantifiers: A Closed Form Formula }
\author{Sagar Malhotra\textsuperscript{\rm 1,2} and Luciano Serafini\textsuperscript{\rm 1}}

 \affiliations{
 \textsuperscript{\rm 1}{Fondazione Bruno Kessler, Italy}\\
 \textsuperscript{\rm 2}{University of Trento, Italy}\\
 smalhotra@fbk.eu, serafini@fbk.eu
 }

\begin{document}

\maketitle

\begin{abstract}
    Weighted First-Order Model Counting (WFOMC) computes the weighted sum of the models of a 
    first-order logic theory on a  given finite domain. 
    First-Order Logic theories that admit polynomial-time WFOMC w.r.t domain cardinality are called domain liftable.
    We introduce the concept of \emph{lifted interpretations} as a tool for formulating closed-forms for WFOMC. Using lifted interpretations, we reconstruct the closed-form formula for polynomial-time FOMC in the universally quantified fragment of FO$^2$, earlier proposed by Beame et al.  
    We then expand this closed-form to incorporate cardinality constraints, existential quantifiers and counting quantifiers (a.k.a. C$^2$) without losing domain-liftability. 
    Finally, we show that the obtained closed-form motivates a natural definition of  a family of weight functions strictly larger 
    than symmetric weight functions.  
  
  \end{abstract}

\section{Introduction}

First-Order Logic (FOL) allows specifying structural knowledge with formulas containing
 variables ranging over all the domain elements.
Probabilistic inference in domains described in FOL requires grounding (aka
instantiation) of all the individual variables with all the occurrences of the
domain elements. This grounding leads to an exponential blow-up of the
complexity of the model description and hence the probabilistic inference.

\emph{Lifted inference} \cite{First_Order_Prob_Inf,de_Salvo} aims at resolving
this problem by exploiting symmetries inherent to the FOL structures. In recent years, \emph{Weighted First-Order Model Counting}
has emerged as a useful formulation for probabilistic inference in statistical relational learning frameworks \cite{SRL_LISA,SRL_LUC}.
Formally, WFOMC \cite{PTP} refers to the task of calculating the weighted sum of the models of
a formula $\Phi$ over a domain of a given finite size 
$$\wfomc(\Phi,w,n) = \sum_{\omega\models\Phi}w(\omega)$$ 
where $n$ is the 
cardinality of the domain and $w$ is a \emph{weight function}
that associates a real number to each interpretation $\omega$. 
FOL theories $\Phi$ and weight functions $w$ which admit polynomial-time WFOMC w.r.t the  domain cardinality  are called
\emph{domain-liftable} \cite{domain_lifted_defintion}. In the past decade, multiple extensions of FO$^2$ (the fragment of
FOL with two variables) have been proven to be
domain-liftable\cite{kazemi2016new,kuusisto2018weighted,kuzelka2020weighted}. 

In this paper, instead of relying on an algorithmic approach to WFOMC,
as in \cite{LIFTED_PI_KB_COMPLETION}, our objective is to find a
closed-form for WFOMC in FO$^2$ that can be easily extended to larger
classes of first-order formulas. To this aim, we introduce the novel
notion of \emph{lifted interpretation}. Lifted interpretations allow us to reconstruct the closed-form
formula for First-Order Model Counting (FOMC) in the universally quantified fragment of FO$^2$ as proposed in \cite{Symmetric_Weighted} and
to extend it to larger classes of FO formulas. 

We see the following key benefits of the presented formulation:
\begin{enumerate}
\item \emph{The formula is easily extended to FO$^2$ with existential quantifiers, cardinality
    constraints and counting quantifiers, without losing domain-liftability}. A cardinality constraint on
  an interpretation is a constraint on the number of elements for
  which a certain predicate holds. Counting quantifiers admit
  expressions of the form $\exists^{\geq m} x\Phi(x)$ 
  expressing that there exist at least $m$ elements that satisfy $\Phi(x)$. Previous 
  works have relied on Lagrange interpolation and Discrete Fourier Transform \cite{kuzelka2020weighted} for evaluating 
  cardinality constraints. In this work, we deal with cardinality constraints in a completely combinatorial
  fashion. 

\item \emph{We provide a complete and uniform treatment of WFOMC in the two-variable fragment}. 
Multiple extensions of FO$^2$ have been proven to be domain liftable \cite{kuzelka2020weighted,kuusisto2018weighted,broeck2013}. Most of these works rely extensively on a variety of  logic-based algorithmic techniques. 
In this paper, we provide a uniform and self-contained combinatorial treatment for all these extensions.  
\item \emph{The formula computes WFOMC for a class of weight functions
    strictly larger than symmetric weight functions.}  The extended
  class of weight functions allows modelling the recently introduced
  count distributions.
\end{enumerate}

Most of the paper focuses on First-Order Model Counting (FOMC) i.e. counting the number of models of a formula $\Phi$ over a finite domain of size $n$ denoted by $\fomc(\Phi,n)$. We
then show how WFOMC can be obtained by multiplying
each term of the resulting formula for FOMC with the corresponding
weight function. This allows us to separate the treatment of the counting part
from the weighting part. The paper is therefore structured as follows:
The next section describes the related work in the literature on
WFOMC. We then present our formulation of the closed-form formula for FOMC
given in \cite{Symmetric_Weighted} for the universally quantified fragment
of FO$^2$. We then extend this formula to incorporate cardinality constraints, existential quantification and counting quantifiers, dedicating one section to each of them respectively. 
The last part of the paper extends
the formula for FOMC to WFOMC for the case of symmetric weight
functions and for a larger class of weight functions that allow modeling 
count distributions \cite{complexMLNkuzelka2020}.

\section{Related Work}
WFOMC for the purposes of probabilistic inference was initially defined and proposed in \cite{PTP} and \cite{LIFTED_PI_KB_COMPLETION}. 
\cite{LIFTED_PI_KB_COMPLETION}  provides an algorithm for Symmetric-WFOMC over universally quantified theories based on \emph{knowledge compilation} techniques.
The notion of a \emph{domain lifted theory} i.e. a first-order theory for which WFOMC can be computed in polynomial time w.r.t domain cardinality was first formalized in \cite{domain_lifted_defintion}. 
The same paper shows that a theory composed of a set of universally quantified clauses containing at most two variables is domain liftable. \cite{broeck2013} extends this procedure to theories in full $\mathrm{FO^{2}}$ (i.e. where existential quantification is allowed) 
  by introducing a  skolemization procedure for WFOMC. 
  
  Theoretical aspects of WFOMC are  analyzed in \cite{Symmetric_Weighted}, which also provides a closed-form formula for WFOMC in the universally quantified fragment of FO$^2$. 
  \cite{kuusisto2018weighted} extends the domain liftability results to FO$^2$ with a functionality axiom, 
  and for sentences in \emph{uniform one-dimensional fragment} U$_1$ \cite{kieronski2015uniform}. 
  It also proposes a closed-form formula for WFOMC in $\mathrm{FO^{2}}$ with functionality constraints. 
  \cite{kuzelka2020weighted} recently proposed a uniform treatment of WFOMC for FO$^2$ with cardinality constraints  and counting quantifiers, 
  proving these theories to be domain-liftable.

  With respect to the state-of-the-art approaches to WFOMC, 
  we propose an approach that provides a closed-form for WFOMC with cardinality constraints and counting quantifiers from which the PTIME data complexity is immediately evident. 
  Moreover, \cite{kuzelka2020weighted} relies on a sequence of reductions for proving 
  domain liftability of counting quantifiers in the two variable fragment, on the other hand, our approach relies on a single reduction and 
  exploits the principle of inclusion-exclusion to provide a closed-form formula for WFOMC. Finally, \cite{complexMLNkuzelka2020} introduces Complex Markov Logic Networks, 
  which use complex-valued weights and allow for full expressivity over a class of distributions called \emph{count distributions}. 
  We show in the last section of the paper that our formalization is complete w.r.t. this class of distributions without using complex-valued weight functions.

  \section{FOMC for Universal Formulas}
  Let $\L$ be a first-order function free language with equality. 
  A \emph{pure universal formula} in $\mathcal{L}$ is a formula of the form $\forall x_1 \dots \forall x_m.\Phi(x_1,\dots,x_m)$, where $X=\{x_1,\dots,x_m\}$ is a set of $m$ distinct variables occurring in $\Phi(x_1,\dots,x_m)$ and $\Phi(x_1,\dots,x_m)$ is a quantifier
  free formula that does not contain any constant symbol. We use the
  compact notation $\Phi(\bx)$ for $\Phi(x_1,\dots,x_m)$, where
  $\bx=(x_1,\dots,x_{m})$. Notice that we distinguish between the
  $m$-tuple of variables $\bx$ and the \emph{set} of variables denoted
  by $X$. We use $C$ to denote the set of domain constants. For every $\bm{\sigma}=(\sigma_1,\dots,\sigma_m)$,
  $m$-tuple of constants or variables, 
  $\Phi(\bm\sigma)$ denotes the
  result of uniform substitution of $x_i$ with $\sigma_i$ in
  $\Phi(\bx)$. 
  If $\Sigma\subseteq X\cup C$ is the set of constants or variables of
  $\L$ and $\forall \bx \Phi(\bx)$ a pure universal formula then:
  \begin{equation}
  \label{Phi_set}
  \Phi(\Sigma) = \bigwedge_{\bm\sigma\in\Sigma^{m}}\Phi(\bm\sigma)
  \end{equation}
  $\Phi(\Sigma)$ is a very convenient notion, for instance, grounding of a pure universal formula $\forall \bx. \Phi(\bx)$ over a set of domain constants $C$, can be simply denoted as 
  $\Phi(C)$. Furthermore, $\Phi(X)$ and $\Phi(\bx)$ have the following useful relationship: 
  \begin{restatable}{lemma}{ccflemma}
    \label{lemm:LI_PERM}
  For any arbitrary pure universal formula  $\forall \bx \Phi(\bx)$, the following equivalence holds: 
  \begin{equation}
      \forall \bx \Phi(\bx) \leftrightarrow \forall \bx  \Phi(X)
  \end{equation}
  \end{restatable}
  
  \begin{example}
    \label{ex:running}
    Let $\Phi(x,y) = A(x) \land R(x,y) \land x\neq y\rightarrow A(y)$,
    then $\Phi(X=\{x,y\})$ is the following formula 
  \begin{align}
  \begin{array}{l@{\ }l} 
  &(A(x) \land R(x,x)\land x\neq x\rightarrow A(x)) \\
  \land &(A(x) \land R(x,y)\land x\neq y  \rightarrow A(y))\\
  \land &(A(y) \land R(y,x)\land y\neq x\rightarrow A(x)) \\
  \land &(A(y) \land R(y,y)\land y\neq y \rightarrow A(y))
  \end{array}
  \label{eq:example}
  \end{align}
  \end{example}
  Due to Lemma \ref{lemm:LI_PERM}, we can assume that in any grounding of $\forall x \forall y.\Phi(X = \{x,y\})$, two distinct variables $x$
  and $y$, are always grounded to different domain elements. This is because the cases
  in which $x$ and $y$ are grounded to the same domain element are taken
  into account by the conjuncts $\Phi(x,x)$ and $\Phi(y,y)$ in $\Phi(X)$. See, for
  instance, the first and the last conjunct of \eqref{eq:example}.
  
  \begin{definition}[Lifted interpretation]
  A \emph{lifted interpretation} $\tau$ of a pure universal formula 
  $\forall\bx\Phi(\bx)$ is a function that assigns to each atom of
  $\Phi(X)$ either $0$ or $1$ ($0$ means false and $1$ means true) and
  assigns 1 to $x_i=x_i$ and 0 to $x_i=x_j$ if $i\neq j$. 
  \end{definition}
  Lifted interpretations are different from FOL interpretations
  as they assign truth values to
  the atoms that contain free variables. Instead, lifted interpretations are similar to 
  $m$-types \cite{kuusisto2018weighted}(we will later formalize this similarity), where $m$ is the number of variables in the language $\mathcal{L}$.
  The \emph{truth value} of a pure universal formula
  $\Phi$ under the lifted interpretation $\tau$ denoted by $\tau(\forall
  \bx\Phi(\bx))$, can be computed by applying the classical semantics for propositional
  connectives to the evaluations of the atoms in $\Phi(X)$. With
  abuse of notation, we sometimes write also $\tau(\Phi(X))$ instead of
  $\tau(\forall \bx\Phi(\bx))$.
  
  \begin{example}
    \label{ex:truth-assignment}
  The following is an  example of a  possible lifted interpretation  $\tau$ for the formula \eqref{eq:example} of Example~\ref{ex:running}:
  $$
  \begin{array}{lcccccc} 
   A(x) & R(x,x) & A(y) & R(y,y) & R(x,y) & R(y,x) \\ \hline
     \cgreen 0 & \cgreen1 & \cred1 & \cred1 & \cgreen0 & \cgreen1 \\
    & \multicolumn{2}{c}{\tau_x} 
         & \multicolumn{2}{c}{\tau_y}
     & \multicolumn{2}{c}{\tau_{xy}} 
   \end{array}
  $$
  We omit the truth assignments of equality atoms since they are fixed for all lifted interpretations. Clearly, $\tau(\eqref{eq:example}) = 0$.
  \end{example}
  As highlighted in the previous example, 
  any lifted interpretation $\tau$ can be decomposed into a set of 
  partial lifted interpretations $\tau_{Y}$ where $Y\subseteq X$.
  Notice that $\tau_Y$ assigns truth value to all the atoms that contain \emph{all} the
  variables $Y$. For instance, in Example \ref{ex:truth-assignment}, 
  $\tau_{\{x,y\}}$ (denoted as $\tau_{xy}$ in the example) assigns to 
  atoms $R(x,y)$ and $R(y,x)$ only and not to the atoms $R(x,x)$ and $A(x)$.
  In general, we will use the simpler notation $\tau_{xy}$ to denote the partial lifted interpretation $\tau_{\{x,y\}}$.

  \cite{Symmetric_Weighted} provide
  a mathematical formula for computing $\fomc(\Phi,n)$, where $\Phi$ is a
  pure universal formula in FO$^2$, i.e., sentences of the form
  $\forall xy.\Phi(x,y)$. In the following, we reconstruct
  this result using the notion of lifted interpretations. As it will be
  clearer later, using lifted interpretations allow us to seamlessly
  extend the result to larger extensions of FO$^2$ formulas.
  
  Let $\forall xy.\Phi(x,y)$ be a pure universal formula. Let $u$ be
  the number of atoms whose truth values are assigned  by $\tau_x$ i.e., first-order atoms containing only the variable $x$. Let $P_1(x),\dots,P_u(x)$ be an ordering of these atoms%
  \footnote{The list includes atoms of the form  $P(x,x)$
    for binary predicate $P$.}. There are $2^u$ possible partial lifted interpretations $\tau_x$ which could assign truth values to these atoms. We assume that each such partial lifted interpretation $\tau_x$ is indexed by an integer $i$, where $0 \leq i \leq 2^u-1$. Hence, the $i$-th partial lifted
  interpretation $\tau_x$ is defined as $\tau_x(P_j(x))=bin(i)_j$, where $bin(i)_j$ represents the value of the $j^{th}$ bit in the binary encoding of $i$, for all $1\leq
  j\leq u$. We use $i(x)$ to denote the conjunction
  of a maximally consistent set of literals (atoms and negated atoms) containing only the
  variable $x$ which are satisfied by the $i$-th $\tau_x$. For instance, in Example \ref{ex:running}, $A(x)$ and $R(x,x)$ are the atoms assigned by $\tau_x$. Assuming the order of atoms to be $(A(x),R(x,x))$, we have that $\tau_x = 1$ implies 
  that $A(x)$ is interpreted to be false and $R(x,x)$ is interpreted to be true. Also, $i(x)$
  denotes $\neg A(x) \land R(x,x)$ if $i=1$ and  $\neg A(x) \land \neg R(x,x)$ if $i=0$. We use similar notation for atoms assigned by  $\tau_y(y)$ i.e. atoms containing only the variable $y$. Furthermore, we use $i(c)$ to denote the conjunction of ground atoms containing only one constant $c$. In Example \ref{ex:running},  $i(c)$ denotes $\neg A(c) \land R(c,c)$ if $i=1$ and  $\neg A(c) \land \neg R(c,c)$ if $i=0$. Clearly, $i(x)$ exactly corresponds to  1-types. Furthermore, given an interpretation $\omega$ if $\omega \models i(c)$ then we say that $c$ is of 1-type $i$.

  For a pure universal formula $\Phi$ and $0\leq i\leq
  j\leq 2^u-1$, let $n_{ij}$ be the number of lifted interpretations $\tau$ that
  satisfy $\Phi$ such that $\tau_x=i$ and $\tau_y=j$. Formally:
  $$
  n_{ij}=\left|\left\{\tau\mid\tau\models\Phi(\{x,y\})\wedge
                                  i(x)\wedge j(y) \right\}\right|
  $$

  \begin{example}[Example \ref{ex:running} cont'd]
    \label{ex:n-ij}
    The set of atoms containing only $x$ or only $y$ in the formula
    \eqref{eq:example} are $\{A(x), R(x,x)\}$ and $\{A(y), R(y,y)\}$
    respectively. In this case $u=2$. The partial lifted interpretations
    $\tau_x$ and $\tau_y$ corresponding to the lifted
      interpretation $\tau$
    of Example~\ref{ex:truth-assignment} are:
    $\tau_x=1$ and $\tau_y=3$. $n_{13}$ is the number of lifted interpretations
    satisfying   \eqref{eq:example} and agreeing with
    $\tau_x=1$ and $\tau_y=3$. In this case $n_{13}=2$.
    The other cases are as follows:
    $$
   \begin{array}{cccccccccc} \hline 
     n_{00} &    n_{01} &    n_{02} &    n_{03} &    n_{11} &    n_{12}
     &    n_{22} &    n_{23} &    n_{33} \\
     4 &    4 &    2 &    2 &    4 &    2 &    4 &    4 &    4 \\ \hline 
   \end{array}
   $$
  \end{example}

  \begin{theorem}[Beame et al. (2015)]
    \label{thm:mc-pure-univ-formulas}
    For any pure universal formula $\forall xy. \Phi(x,y)$ 
    \begin{align}
      \label{eq:fomc-universal-form}
      \fomc(\forall xy. \Phi(x,y),n) = &
     \sum_{\sum \bk=n}  \binom{n}{\bm{k}}
     \prod_{0\leq i\leq j\leq 2^u-1}\!\!\! n_{i j}^{\bk(i,j)}
  \end{align}
  where $\bk = (k_0,...,k_{2^{u}-1})$ is a $2^{u}$-tuple of non-negative integers, $\binom{n}{\bm{k}}$ is the multinomial coefficient and
  $$
    \bk(i,j) =
    \begin{cases}
        \frac{k_{i}(k_{i} - 1)}{2} & \text{if $i=j$} \\
         k_{i}k_{j} & \text{otherwise} \\
       \end{cases}
  $$
  \end{theorem}
  
  We provide the proof for Theorem \ref{thm:mc-pure-univ-formulas} along with some additional  Lemmas in the appendix. Intuitively, $k_i$ represents the number of constants $c$ of 1-type $i$.
  Hence, for a given 
  $\bk$, we have $\binom{n}{\bk}$ possible interpretations. Furthermore, given a pair of constants $c$ and $d$ such that 
   $c$ is of 1-type $i$ and $d$ is of 1-type $j$, the number of extensions to the binary predicates containing both $c$ and $d$ such that the extensions are  models of $\forall xy.\Phi(x, y)$, is given by $n_{ij}$ independently of all other constants. Finally, the exponent $\bk(i,j)$ accounts for all possible pair-wise choices of constants given a $\bk$ vector.

  Notice that the  method in \cite{Symmetric_Weighted} requires
  additional $n+1$ calls to a counting oracle for dealing with equality.  Lifted interpretations on the other hand allow us to fix the truth values of the equality atoms, by assuming (w.l.o.g.) that different variables are assigned to distinct objects in $\forall \bx.\Phi(X)$. The equality atoms then contribute to the model count only through $n_{ij}$, hence, allowing us to deal with equality in constant time w.r.t domain cardinality. 
  
  \begin{example}[Example \ref{ex:running} continued]
    \label{ex:cardinality}
    Consider a domain of 3 elements (i.e., $n=3$). Each term of the summation
    \eqref{eq:fomc-universal-form} is of the form
    $$
    \binom{3}{k_0,k_1,k_2,k_3}\prod_{i=0}^3n_{ii}^{\frac{k_i(k_i-1)}{2}}\prod_{\substack{i<j \\i=0 }}^{3}n_{ij}^{k_ik_j}
    $$
    which is the number of models with $k_0$ elements for which $A(x)$
    and $R(x,x)$ are both false; $k_1$ elements for which $A(x)$ is false
    and $R(x,x)$ true, $k_2$ elements for which $A(x)$ is true and
    $R(x,x)$ is false and $k_3$ elements for which $A(x)$ and
    $R(x,x)$ are both true. For instance:
    $
    \binom{3}{2,0,0,1} n_{00}^1n_{03}^2= \binom{3}{2,0,0,1} 4^1\cdot 2^2
    = 3\cdot 16 = 48
    $ is the number of models in which 2 elements are such that $A(x)$ and
    $R(x,x)$ are false and $1$ element such that $A(x)$ and $R(x,x)$ are both
    true. 
    \end{example}
  
    \section{FOMC for Cardinality Constraints}
    Cardinality constraints are arithmetic expressions that impose
    restrictions on the number of times a certain predicate is interpreted to be true. A
    simple example of a cardinality constraint is $\card{A}=m$, for some
    unary predicate $A$ and positive integer $m$. This cardinality constraint is
    satisfied by any interpretation in which $A(c)$ is interpreted 
    to be true for exactly $m$ distinct constants $c$ in the domain $C$.
    A more complex example of a cardinality constraint could be: 
    $\card{A}+\card{B} \leq |C|$, where $A$, $B$ and $C$ are some predicates in the 
    language.

    For every interpretation $\omega$ of the
    language $\L$ on a finite domain $C$, we define
    $A^\omega=\{c\in C\mid \omega\models A(c)\}$ if $A$ is unary, and
    $A^\omega=\{(c,d)\in C\times C\mid \omega\models A(c,d)\}$ if $A$ is
    binary.  $\omega$ satisfies a cardinality constraint $\rho$, in
    symbols $\omega\models\rho$, if the arithmetic expression, 
    obtained by replacing $\card{A}$ with $\card{A^\omega}$ for every predicate $A$ in $\rho$, is satisfied.
    
    If a cardinality constraint involves only unary predicates, then we
    can exploit Theorem~\ref{thm:mc-pure-univ-formulas}
    considering only a subset of $\bk$'s. The multinomial 
    coefficient $\binom{n}{\bk}$ counts the models that contain
    exactly $k_i$ elements of $1$-type $i$, the cardinality of the unary
    predicates in these models are fully determined by $\bk$. 
    
    To deal with cardinality constraints involving binary
    predicates, we have to expand the formula
    \eqref{eq:fomc-universal-form} by including also the assignments to
    binary predicates.  This implies extending the $\bk$ vector in order
    to consider assignments to atoms that contain both
    variables $x$ and $y$.
    Let $R_0(x,y),R_1(x,y),\dots,R_b(x,y)$ be an enumeration of all the atoms
    in $\Phi(X)$ that contain both variables $x$ and $y$. Notice that the
    order of variables leads to different atoms, for instance in Example
    \ref{ex:running},  we have two binary atoms $R_1(x,y)=R(x,y)$ and
    $R_2(x,y)=R(y,x)$. 
    
    For every $0 \leq v\leq 2^b-1$ let $v$ denote the
    $v^{th}$ partial lifted interpretation $\tau_{xy}$, such that $\tau_{xy}$ assigns $bin(v)_j$ to
    the $j$-th binary atom $R_{j}(x,y)$ for every $1\leq j\leq b$.
    As for the unary case, $v(x,y)$ represents the conjunction of all the
    literals that are satisfied by $v$. For instance, in example \ref{ex:running}, $v(x,y)$
    denotes $\neg R(x,y) \land R(y,x)$ if $v = 1$ and $R(x,y) \land  R(y,x)$ if $v=3$. Clearly, the set of 2-types in the language of the formula $\Phi$ correspond to $i(x)\land j(y) \land v(x,y)$. We define $n_{ijv}$ as follows: 
    $$
    n_{ijv}=\left|\left\{\tau\mid\tau\models\Phi(\{x,y\})\wedge
                                    i(x)\wedge j(y)\wedge v(x,y)\right\}\right|
    $$
   
    Notice that $n_{ij}=\sum_{v=0}^{2^{b}-1}n_{ijv}$ and that $n_{ijv}$ is either $0$
    or $1$. 

    \begin{example}
      For instance $n_{13}$ introduced in Example~\ref{ex:n-ij}
      expands to $n_{130}+n_{131}+n_{132}+n_{133}$
      where $n_{13v}$ corresponds to the following assignments:
      $$
    \arraycolsep=2pt
      \begin{array}{|cc;{1pt/1pt}cc;{1pt/1pt}cc|c|} \hline 
     A(x) & R(x,x) & A(y) & R(y,y) & R(x,y) & R(y,x) & n_{13v}\\ \hline
        \multirow{4}{*}{$0$} & 
        \multirow{4}{*}{$1$} & 
        \multirow{4}{*}{$1$} & 
        \multirow{4}{*}{$1$} & 
        0 & 0 & n_{130} = 1 \\ 
        & & & & 0 & 1 & n_{131}=0\\ 
        & & & & 1 & 0 & n_{132}=1\\ 
        & & & & 1 & 1 & n_{133}=0\\ \hline 
        \multicolumn{2}{c}{\tau_x=1} 
           & \multicolumn{2}{c}{\tau_y=3}
       & \multicolumn{2}{c}{\tau_{xy}=v} \\ 
      \end{array}
    $$
    
    \end{example}

     By replacing $n_{ij}$ in equation \eqref{eq:fomc-universal-form} with
     its expansion $\sum_{v=0}^{2^{b}-1}n_{ijv}$ we obtain that
    $\fomc(\forall xy. \Phi(x,y),n)$ is equal to 
    
\begin{align}
    &  \sum_{\sum\bk=n}  \binom{n}{\bm{k}}
      \prod_{0 \leq i \leq j \leq 2^u-1} \left(\sum_{0\leq v \leq
                            2^b-1}n_{i j v}\right)^{\bk(i,j)} \nonumber \\
                       & = \sum_{\bk,\bh}\binom{n}{\bk}
    \prod_{0 \leq i \leq j \leq 2^u-1}
    \binom{\bk(i,j)}{\bh^{ij}}
    \prod_{0\leq v\leq 2^b-1} 
        n_{ijv}^{h^{ij}_{v}}
    \end{align}
    where, for every $0\leq i\leq j\leq 2^u-1$, $\bh^{ij}$ is a vector of
    $2^b$ integers that sum up to $\bk(i,j)$. To simplify the notation we
    define the function $F(\kh,\Phi)$ where $\Phi$ is a pure universal
    formula as follows 
    $$
    F(\kh,\Phi)= \binom{n}{\bk}\!\!
    \prod_{0 \leq i \leq j \leq 2^u-1}\!\!
    \!\!\binom{\bk(i,j)}{\bh^{ij}}\!\!
    \prod_{0\leq v\leq 2^b-1} \!\!\!\!
        n_{ijv}^{h^{ij}_{v}}
        $$
    where $h^{ij}_{v}$ is the $v$-th element of the vector $\bh^{ij}$, which  represents
    the number of pairs of constants of distinct elements that satisfy the 2-type $i(x)\wedge
    j(y)\wedge v(x,y)$. We will now show that the $(\kh)$ vectors contain
    all the necessary information for determining the cardinality of the binary predicates.
    
    For every $\L$-interpretation $\omega$ on the finite domain $C$, we define
    $(\kh)^\omega=(\bk^\omega,\bh^\omega)$ with 
    $\bk^{\omega}=\left<k^\omega_0,\dots,k^\omega_{2^u-1}\right>$ such
    that $k_i^\omega$ is the number of constants $c\in C$ such that
    $\omega\models i(c)$.
    $\bh^\omega$ is equal to $\{(\bh^{ij})^\omega\}_{0\leq i\leq j\leq 2^b-1}$, where 
    $(\bh^{ij})^\omega=\langle(h_0^{ij})^\omega,\dots
      (h_{2^b-1}^{ij})^\omega\rangle$ such that $(h_v^{ij})^\omega$ is the
    number of pairs $(c,d)$ with $c\neq d$ such that $\omega\models
    i(c)\wedge j(d)\wedge v(c,d)$ if $i < j$. When  $i=j$, $(h^{ii}_v)^{\omega}$ is
    equal to the count of the unordered pairs $(c,d)$ (i.e. only one of
    the $(c,d)$ and $(d,c)$ is counted) for which $\omega \models i(c)\land i(d) \land v(c,d)$.
    \begin{lemma}
      \label{lem:cardinality}
      For every predicate $P$ and interpretations $\omega_1$ and $\omega_2$, $(\kh)^{\omega_1} =(\kh)^{\omega_2}$ implies
      $\card{P^{\omega_1}}=\card{P^{\omega_2}}$. 
    \end{lemma}
    
    \begin{proof}
       Let $(\kh)$ be a vector such that   $(\kh) = (\kh)^\omega$. 
      The Lemma is true iff $(\kh)$ uniquely determines the cardinality of $P^{\omega}$. If $P^{\omega}$ is a unary predicate whose atom is indexed by $s$ in the ordering of the unary atoms, then the cardinality of 
      $P^{\omega}$ can be given as $\sum_{i=0}^{2^{u}-1}bin(i)_s \cdot k_i$. Similarly, if $P$ is binary then in order to count $P^{\omega}$, we need to take into account both
      $\bk$ and $\bh$. Let $P(x,x)$ be the atom indexed $s$ i.e. $P_s$, let $P(x,y)$ be the atom indexed $l$ i.e. $P_l$ and let $P(y,x)$ be the atom indexed $r$ i.e. $P_r$, then the cardinality of 
      $P$ if $P$ is binary is given as $\sum_{i=0}^{2^{u}-1}bin(i)_s \cdot k_i + \sum_{i\leq j}\sum_{v=0}^{2^b-1}(bin(v)_l+bin(v)_r)\cdot h^{ij}_{v}$.

    \end{proof}
    
    \begin{example}
      Consider formula \eqref{eq:example} with the additional conjunct
      $\card{A}= 2$ and $\card{R}=2$. The constraint $\card{A}=2$ implies that we have to
      consider $\bk$ such that $k_2+k_3=2$. $\card{R}=2$ constraint translates to only
      considering $(\kh)$ with
      $k_1+k_3+ \sum_{i\leq j} (h^{ij}_{1}+ h^{ij}_{2} + 2h^{ij}_{3}) = 2$. 
    \end{example}
    
    For a given $(\kh)$, we use the notation $\bk(P)$ to denote cardinality of $P$ if $P$ is unary and $(\kh)(P)$ if $P$ is binary.
    Using Lemma \ref{lem:cardinality}, we can conclude that $\fomc(\Phi\land\rho,n)$ where
    $\Phi$ is a pure universal formula with 2 variables can
    be computed by considering only the $(\kh)$'s
    that satisfy $\rho$, i.e., those $(\kh)'s$ where $\rho$ evaluates to true, when  
    $|P|$ is substituted with $(\kh)(P)$ when $P$ is binary and $\bk(P)$ when $P$ is unary. 
 
    \begin{corollary}[of Theorem~\ref{thm:mc-pure-univ-formulas}]
      For every pure universal formula $\Phi$ and cardinality constraint
      $\rho$,
    $\fomc(\Phi\land \rho, n)    
        =\sum_{\kh\models\rho}F(\kh,\Phi)
    $
 
    \end{corollary}
    \section{FOMC for Existential Quantifiers}
    In this section, we provide a proof for model counting in the presence of existential quantifiers. The key difference in our approach w.r.t
    \cite{Symmetric_Weighted} is that we make explicit use of the principle of inclusion-exclusion, and we will later generalize the same approach to counting quantifiers.  We will first provide a corollary of the principle of inclusion-exclusion. 
    
    \begin{corollary}[\cite{GF_book} section 4.2]
      \label{thm:inclusion-exclusion}
      Let $\Omega$ be a set of objects and let
      $\mathcal{S}=\{S_1,\dots,S_m\}$ be a set of subsets of $\Omega$. For every $\mathcal{Q} \subseteq \mathcal{S}$, let
      $N(\supseteq \mathcal{Q})$ be the count of objects in $\Omega$
      that belong to all the subsets $S_i \in \mathcal{Q}$,
      i.e., $N(\supseteq \mathcal{Q})=\left|\{\bigcap_{S_i\in Q}S_i\}\right|$.
      For every $0\leq l\leq m$, let
      $s_l = \sum_{|\mathcal{Q}|=l}N(\supseteq \mathcal{Q})$ and
      let $e_0$ be count of objects that do not belong to any of the $S_i$ in $\mathcal{S}$, then 
      \begin{equation}
         e_0 = \sum_{l=0}^m (-1)^{l} s_l 
      \end{equation}
    \end{corollary}

    Any arbitrary formula in FO$^2$ can be reduced to an equisatisfiable
    reduction called Scott's Normal Form (SNF) \cite{Scott1962}. Moreover, SNF preserves 
    FOMC as well as WFOMC if all the new predicates and their negation are assigned a unit weight \cite{kuusisto2018weighted}. A formula in SNF has the following form:
        \begin{align}
          \label{eq:simple-scott}
        \forall x \forall y.\Phi(x,y) \land \bigwedge_{i=1}^{q} \forall x \exists y.\Psi_i(x,y)
        \end{align}
        where $\Phi(x,y)$ and $\Psi_i(x,y)$ are quantifier-free formulae.
    
        \begin{theorem}
          \label{thm:fomc-scott-form}
          For an FO$^2$ formula in Scott's Normal Form as given in \eqref{eq:simple-scott}, let $\Phi'= \forall xy.(\Phi(x,y)\wedge\bigwedge_{i=1}^qP_i(x)\rightarrow\neg\Psi_i(x,y))$
          where $P_i$'s are fresh unary predicates, then: 
          \begin{align}
          \label{eq:fomc-scott-form}
          \fomc(\eqref{eq:simple-scott},n) = \!\!
            \sum_{\kh} (-1)^{\sum_i \bk(P_i)}F(\kh,\Phi')
          \end{align}
          \end{theorem}
      
      \begin{proof}\def\bmm{\bm m}
        Let $\Omega$ be the set of models of $\forall xy.\Phi(x,y)$ over the language of $\Phi$ and $\{\Psi_i\}$ (i.e., the language of $\Phi'$ excluding the predicates $P_i$) and on a domain $C$ consisting of $n$ elements. 
        Let $\mathcal{S}=\{\Omega_{ci}\}_{c\in  C,\  1\leq i\leq q}$ be
        the set of subsets of $\Omega$ where $\Omega_{ci}$ is the set of
        $\omega$ such that $\omega\models \forall y. \neg \Psi_i(c,y)$. For every model $\omega$ of \eqref{eq:simple-scott}, 
        $\omega\not\models\forall y\neg \Psi_i(c,y)$ for any pair of $i$ and $c$ i.e.
        $\omega$ is not in any $\Omega_{ci}$. Also, for every $\omega \in \Omega$,  if $\omega \not\in \Omega_{ci}$ for any pair of $i$ and $c$, then $\omega \models \exists y. \Psi_i(c,y)$ for all $i$ and for all $c \in C$ i.e., $\omega \models \bigwedge_{i=1}^{q} \forall x \exists y. \Psi_i (x,y)$. Hence, $\omega \models \eqref{eq:simple-scott}$ if and only if $\omega \not\in \Omega_{ci}$ for all $c$ and $i$.
        Therefore, the count of models of
        \eqref{eq:simple-scott} is equal to the count of models in
        $\Omega$ which do not belong to any $\Omega_{ci}$. Hence, If we are able to compute $s_l$ (as introduced in Corollary \ref{thm:inclusion-exclusion}), then we could use Corollary~\ref{thm:inclusion-exclusion} for computing cardinality of all the models which do not belong to any $\Omega_{ci}$ and hence $\fomc(\eqref{eq:simple-scott},n)$. 
        
        For every $0\leq l\leq n\cdot q$, let us define 
            \begin{align}
              \label{eq:phi-l}
          \Phi'_{l}&= \Phi'\wedge\sum_{i=1}^q|P_i|=l
        \end{align}
        We will now show that $s_l$ is exactly given by $\fomc(\eqref{eq:phi-l},n)$. 
        
        Every model of $\Phi'_{l}$ is an extension of an $\omega \in \Omega$
        that belongs to at least $l$ elements in $\mathcal{S}$. In fact, for every model $\omega$ of $\forall xy.\Phi(x,y)$ i.e. $\omega \in \Omega$, if $\mathcal{Q'}$ is the set of
        elements of $\mathcal{S}$ that contain $\omega$, then 
        $\omega$ can be extended into a model of $\Phi'_l$ in $\binom{|Q'|}{l}$ ways.
        Each such model can be obtained  by choosing $l$ elements in $Q'$ and interpreting $P_i(c)$ to be true in the extended model, for each of the $l$ chosen elements $\Omega_{ci} \in Q'$. On the other hand, recall that  $s_l= \sum_{|\mathcal{Q}|=l} N(\supseteq Q)$. Hence, for any $\omega \in \Omega$  if $\mathcal{Q'}$ is the set of elements of $\mathcal{S}$ that contain $\omega$, then there are $\binom{|\mathcal{Q}'|}{l}$ distinct subsets $\mathcal{Q}\subseteq \mathcal{Q'}$ such that $|\mathcal{Q}|=l$. Hence, we have that $\omega$
        contributes $\binom{|\mathcal{Q}'|}{l}$ times to 
        $s_l$.
        Therefore, we can conclude that
        $$
        s_l=\fomc(\Phi'_l,n)=\sum_{|\mathcal{Q}|=l}N(\supseteq Q)
        $$
      and by the principle of inclusion-exclusion as given in Corollary
      ~\ref{thm:inclusion-exclusion}, we have that :
      \begin{align*}
       \fomc(\eqref{eq:simple-scott},n)  & = e_0 =  \sum_{l=0}^{n\cdot    q}(-1)^l s_l \\
       & =  \sum_{l=0}^{n\cdot    q}(-1)^l\fomc(\Phi'_l,n) \\
       & = \sum_{l=0}^{n\cdot    q}(-1)^l\sum_{\kh\models\sum_i|P_i|=l}F(\kh,\Phi')\\
       & = \sum_{\kh}(-1)^{\sum_i\bk(P_i)}F(\kh,\Phi')
      \end{align*}
    \end{proof}   

    \section{FOMC for Counting Quantifiers}
    \emph{Counting quantifiers} are expressions of the form
    $\exists x^{\geq m}y.\Psi$, $\exists^{\leq m}y.\Psi$, and
    $\exists^{= m}y.\Psi$. The extension of FO$^2$ with such quantifiers
    is denoted by C$^2$ \cite{COUNTING_REF}.  In this section, we show how
    FOMC in C$^2$ can be performed by exploiting the formula for FOMC in
    FO$^2$ with cardinality constraints.
    We assume that the counting
    quantifier $\exists^{\leq m}y.\Psi$ is expanded to
    $\bigvee_{k=0}^m\exists^{=k}y.\Psi$, and the quantifiers 
    $\exists^{\geq m}y.\Psi$ are first transformed to
    $\neg(\exists^{\leq m-1}y.\Psi)$ and then expanded. We are therefore
    left with quantifiers of the form $\exists^{=m}y.\Psi$. Hence, any C$^2$ formula can 
    be transformed into a formula of the form  $ \Phi_0 \land \bigwedge^{q}_{k=1} \forall x.(A_{k}(x) \leftrightarrow
    \exists^{=m_k}y. \Psi_{k} )$ that preserves FOMC, where\footnote{We assume that $\Phi_0$ contains no existential quantifiers as they can be transformed as described in Theorem \ref{thm:fomc-scott-form}.} $\Phi_0$ is a pure universal formula obtained by replacing every occurrence of the sub-formula $\exists^{= m_k}y.\Psi_k$ with $A_{k}(x)$, where $A_k$ is a fresh predicate. W.l.o.g, we can assume that $\Psi_k$ is the atomic formula $R_k(x,y)$. We will now present a closed-form for FOMC of
    $\Phi_0 \land \bigwedge_{k} \forall x. (A_{k}(x) \leftrightarrow \exists^{=m_{k}}y.R_{k}(x,y))$. For the sake of notational convenience, we use $\Phi_{i..j}$ to denote $\bigwedge_{i\leq s \leq j} \Phi_s$ for any set of formulas $\{\Phi_s\}$.

    \begin{theorem}
      Let $\Phi$ be the following C$^2$  formula :
      $$\Phi_0 \land \bigwedge^{q}_{k=1} \forall x.(A_{k}(x) \leftrightarrow
      \exists^{=m_k}y. R_{k}(x,y) )$$
      where $\Phi_0$ is a pure universal formula
      in FO$^2$. Let us define the following formulas for each $k$, where $1\leq k \leq q$:
      \label{thm:Counting1}
      \begin{align*}
          \Phi^{k}_{1} &= \mbox{$\bigwedge_{i=1}^{m_k}$} \forall x \exists y.A_{k}(x)\lor B_{k}(x)\rightarrow  f_{ki}(x,y)\\
          \Phi^{k}_{2} &= \mbox{$\bigwedge_{1\leq i< j\leq m_k}$}\forall x \forall y.f_{ki}(x,y) \rightarrow\neg f_{kj}(x,y)\\
          \Phi^{k}_{3} &=\mbox{$\bigwedge_{i=1}^{m_k}$} \forall x \forall y. f_{ki}(x,y) \rightarrow R_{k}(x,y) \\
          \Phi^{k}_4 &= \forall x. B_{k}(x) \rightarrow \neg A_{k}(x)\\
          \Phi^{k}_5 &= \forall x \forall y. M_{k}(x,y) \leftrightarrow ((A_{k}(x)\lor B_{k}(x))\land R_{k}(x,y)) \\
        \Phi^{k}_6 &= \card{A_{k}}+\card{B_{k}} = \card{f_{k1}}=\dots=\card{f_{km_k}} = \mbox{$\frac{\card{M_{k}}}{m_{k}}$}
      \end{align*}
      where\footnote{If $\Phi_0$ is obtained after a transformation as described in Theorem \ref{thm:fomc-scott-form}, then we can add the term 
    $\sum_{g}k(P_g)$ to the exponent of $(-1)$, for the set of unary predicates $\{P_g\}$ introduced to deal with existential quantifiers. Also, any cardinality constraint on predicates of $\Phi_0$ can be easily conjuncted and incorporated into  $\land_k\Phi^{k}_6$.} $B_{k}$, $f_{ki}$ and $M_{k}$ are fresh predicates. Then $\fomc(\Phi,n)$ is given as:
    $$
    \sum_{(\kh)\models \bigwedge_k \Phi^{k}_6}\frac{(-1)^{\sum_{k}\bk(B_k)+\sum_{k,i}\bk(P_{ki})} F(\kh,\Phi')}{\prod_k m_k!^{\bk(A_k)}}
    $$
    where $\Phi'$ is obtained by replacing each $\Phi^{k}_1$ with   $ \mbox{$\bigwedge_{i=1}^{m_k}$} \forall x \forall y. P_{ki}(x) \rightarrow \neg(A_{k}(x)\lor B_{k}(x)\rightarrow f_{ki}(x,y))$ 
    in $\Phi_0 \land \bigwedge_k \Phi^{k}_{1..5}$ and $P_{ki}$ are fresh unary predicates.

    \end{theorem}
    \begin{lemma}
    \label{lem:counting}
    If $\omega\models \Phi_0 \land \bigwedge_{k=1}^{q}\Phi^{k}_{1..6}$
    then every $c\in A^\omega_k\cup B^\omega_k$ has exactly $m_k$
    $R_k$-successors i.e., $\omega\models\exists^{=m_k}y.R(c,y)$.
    \end{lemma}
    
    \begin{proof}
      If $c\in A^\omega_k\cup B^\omega_k$, then by $\Phi^k_1$, $c$ has an 
      $f_{ki}$-successor for every $1\leq i\leq m_k$.
      $\Phi^{k}_2$ implies that $c$ has distinct  $f_{ki}$ and $f_{kj}$ successor for 
      any choice of $i$ and $j$. $\Phi^{k}_3$ implies that any $f_{ki}$-successor of $c$ is also an $R_k$-successor. 
      Hence, $c$ has at least $m_k$ $R_k$-successors.

      Axiom $\Phi^{k}_5$ implies that $c$ has exactly as many $R_k$-successors as $M_k$-successors.
      Hence, $c$ has at-least $m_k$ $M_k$-successors. Furthermore, by $\Phi^{k}_4$ we have that $A_{k}^{\omega}$ and $B_{k}^{\omega}$ are disjoint. 
      Hence, using $\Phi^{k}_6$, we can conclude that $c$ has exactly $m_k$ $M_k$-successors. Finally, using $\Phi^{k}_5$ we can conclude that $c$ has
      exactly $m_k$ $R_k$-successors.
    \end{proof}
    \begin{proof}[Proof (of Theorem \ref{thm:Counting1})]
      First notice that every model $\omega$ of $\Phi$ can be extended to $\prod_km_k!^{A_k^\omega}$ models of
      $\Phi_0\wedge\bigwedge_k\Phi^k_{1..6}$ by interpreting $B_k$ in the
      empty set, 
      $f_{ki}$ in the set of pairs $\left<c,d\right>$ for $c\in A_k^\omega$ and $d$
        being the $i$-th $R_k$-successor of $c$ (for some ordering of the
        $R_k$-successors) and $M_{k}$ according to the
        definition given in $\Phi^5_k$.

Let $\Omega$ the set of models of $\Phi_0 \land \bigwedge_{k=1}^{q} \Phi^{k}_{1..6}$ restricted to the
      language of $\Phi$, $M_k$ and $f_{ki}$ (i.e., the language of $\Phi_0 \land \bigwedge_{k=1}^{q} \Phi^{k}_{1..6}$  excluding the predicates $B_k$) and on a domain $C$ consisting of $n$ elements. 
      
      Notice that $\Omega$ contains also the models that are not extensions of
      some model of $\Phi$. Therefore, in the first part of the proof we count
      the number of extensions of models of $\Phi$ in $\Omega$, and
      successively we will take care of the over-counting due to the
      multiple interpretations of $f_{ki}$'s. 
      
      Let 
      $\mathcal{S}=\{\Omega_{ck}\}$ be the set of subsets of
      $\Omega$ such that if $\omega \in \Omega_{ck}$
     then $\omega \models \neg A_k(c)\wedge\exists^{=m_k}y.R_k(c,y)$.
      Due to Lemma \ref{lem:counting}, if $\omega \in \Omega$  
      then $\omega \models \bigwedge_{k} \forall x.A_k(x) \rightarrow \exists^{=m_k}y.R_k(x,y)$.
      Hence, in order to count the models of $\Phi$ in $\Omega$ we only need to count the number of models in $\Omega$ that
      satisfy $\bigwedge_{k}\forall x\exists^{=m_k}y.R_k(x,y)\rightarrow A_k(x)$, equivalently, the number of models that belong to 
      none of the $\Omega_{ck}$. Hence, if we are able to evaluate $s_l$ (as introduced in Corollary \ref{thm:inclusion-exclusion}) then we can use Corollary $\ref{thm:inclusion-exclusion}$ to
      count the set of models in $\Omega$ that satisfy $\Phi$. 
      
      Let $\omega\in\Omega$. Let us define $\Phi_{l}$  for $l \geq 0$ as follows:
      \begin{equation}
      \label{eq:Phi_l_count_1}  
      \Phi_{l} = \Phi_0 \land \bigwedge_k\Phi^{k}_{1..6}\wedge \left(\sum_k\card{B_k}=l\right)
      \end{equation}
      Firstly, let $\mathcal{Q'}$ be the set of elements in $\mathcal{S}$ that contain $\omega$.
      By Lemma~\ref{lem:counting}, 
      $\omega$ can be extended in $\binom{|\mathcal{Q'}|}{l}$ models of
      $\Phi_l$. Each such extension can be achieved by choosing 
      $l$ elements in $\mathcal{Q'}$, and interpreting $B_k(c)$ to be true in the extended model iff 
      $\Omega_{ck}$ is a part of the $l$ chosen elements. On the other hand, recall that 
      $s_l=\sum_{|\mathcal{Q}|=l}N(\supseteq\mathcal{Q})$. Every $\omega$ that is contained
      in all the elements of $\mathcal{Q'}$, contributes $\binom{|\mathcal{Q'}|}{l}$ to $s_l$.
      Hence,
      $
      s_l = \fomc(\Phi_l,n)
      $.
      Using inclusion-exclusion principle (corollary \ref{thm:inclusion-exclusion}), we have that
      the number of models which do not belong to any of the $\Omega_{ck}$ are:
      \begin{align}
        \label{eq:proof-counting-1}
     &\sum_{l}(-1)^ls_l = \sum_{l}(-1)^l\fomc(\Phi_l,n)
      \end{align}
      Hence, we have the count of models of $\Phi$ in $\Omega$.
      But notice that this is the count of the models of $\Phi$ in the language of
      $\Phi_0 \land \bigwedge_k \Phi^{k}_{1..6}$ excluding $B_k$, where there are the additional
      predicates $\{f_{ki}\}$. Since every interpretation with $|A_k^\omega|=r_k$ 
    can be extended in $m_k!^{r_k}$ models of $\Phi$ due to the permutations of $\{f_{ki}\}_{i=1}^{m_k}$, to obtain FOMC on the 
    language of $\Phi$ we have to take into account this over-counting%
    \footnote{Notice that $M_k$ leads to 
    no additional models of $\Phi$ as interpretations of $M_k$ are
    uniquely determined by $A_k$ and $R_k$ by $\Phi^k_5$.}.
    This can be obtained by introducing a cardinality constraint $\card{A_k}=r_k$ for every $A_k$
    and dividing  by $m_k!^{r_k}$ for each $k$ and $r_1...r_q$ values. Giving the following expression for 
     $\fomc(\Phi,n)$: 
      \begin{align}
        \label{eq:proof-counting-2}    
      \sum_{l,r_k}(-1)^l\frac{\fomc(\Phi_{l}\land\bigwedge_{k}|A_k|=r_k,n)}{\prod_k m_k!^{r_k}}
    \end{align}
    Also notice that $\Phi^{k}_1$ contains $m_k$ existential quantifiers, 
    to eliminate them we use the result of
    Theorem~\ref{thm:fomc-scott-form}.
    We introduce $m_k$ new unary
    predicates $P_{k1},\dots,P_{km_k}$ for each $k$, and replace each $\Phi^{k}_1$ with $\bigwedge_i \forall x \forall y. P_{ki}(x) \rightarrow \neg(A_k(x) \lor B_k(x) \rightarrow f_{ki}(x,y))$. Hence, by Theorem~\ref{thm:fomc-scott-form} we have that $\fomc(\Phi,n)$ is equal to: 
    $$
    \sum_{(\kh)\models \bigwedge_{k}\Phi^{k}_6}\frac{(-1)^{\sum_{k}\bk(B_k)+\sum_{k,i}\bk(P_{ki})} F(\kh,\Phi')}{\prod_k m_k!^{\bk(A_k)}}
    $$
    where $\Phi'$ is the pure universal formula
    $\Phi_0 \land \bigwedge_{k=1}^{q}\Phi^{k}_{2..5} \land \bigwedge_{i,k}
    P_{ki}(x) \rightarrow \neg(A_{k}(x)\lor B_{k}(x)\rightarrow
    f_{ki}(x,y))$.
  
  \end{proof}

    \def\bc{\bm{c}}
    \def\wfomc{\text{\footnotesize\sc wfomc}}
    \section{Weighted First-Order Model Counting}
    All the FOMC formulas introduced so far  can be easily extended to weighted 
    model counting by simply defining a positive real-valued weight function $w(\kh)$ and adding it as a multiplicative 
    factor to $F(\kh,\Phi)$ in all FOMC formulas. The case of Symmetric-WFOMC can be obtained by defining $w(\kh)$ as follows:
    
    \begin{align*}
       w(\bk,\bh) & = \prod_{P\in\L}
       w(P)^{(\bk,\bh)(P)}\cdot 
       \bar{w}(P)^{(\bk,\bh)(\neg P)}
    \end{align*}
    where $w(P)$ and $\bar{w}(P)$ associate positive real values to predicate $P$ and its negation respectively. But symmetric-weight functions are  clearly not the most general class of weight functions.
    \cite{complexMLNkuzelka2020} introduced a strictly more expressive class of weight functions
    which also preserves domain liftability. These weight functions can express count distributions, which are defined as follows:
    
    \begin{definition}[Count distribution \cite{complexMLNkuzelka2020}]
    Let $\Phi = \{\alpha_i, w_i\}_{i=1}^m$
    be a Markov Logic Network defining a probability distribution
    $p_{\Phi,\Omega}$ over a set of
    possible worlds (we call them assignments) of a formula
    $\Omega$. The count distribution of $\Phi$ is the distribution
    over $m$-dimensional vectors of non-negative integers $\bm n$ given by
    \begin{align}
    q_{\Phi}(\Omega,\bm n) & = \sum_{\omega\models\Omega,\ \bm n =
                       \bm N(\Phi,\omega)} p_{\Phi,\Omega}(\omega)
    \end{align}
    where $\bm N(\Phi,\omega)=(n_1,\dots,n_m)$ and $n_i$ is the number of
    grounding of $\alpha_i$ that are true in $\omega$. 
    
    \end{definition}

    \cite{complexMLNkuzelka2020} shows that count distributions can
    be modelled by Markov Logic Networks with complex weights. In the following, we prove
    that if each $\alpha_i$ is in FO$^2$, count distributions
    can be expressed by a $w(\kh)$. 
    
    \begin{theorem}
      \label{thm:Counting_WFOMC}
    Every count distribution over a set of possible worlds of a formula $\Omega$ 
    definable in FO$^2$ can be modelled with a weight
    function on $(\kh)$, by introducing $m$ new predicates $P_i$ 
    and adding the axioms $P_i(x)\leftrightarrow \alpha_i(x)$ and 
    $P_j(x,y)\leftrightarrow\alpha_j(x,y)$, if $\alpha_i$ and $\alpha_j$
    has one and two free variables respectively and by defining: 
    \begin{align}
     \label{eq:conting-distribution}
    q_{\Phi}(\Omega,\bm n) & =\frac{1}{Z}
    \sum_{(\bk,\bh)(P_i)=n_i}w(\bk,\bh)\cdot F(\bk,\bh,\Omega)
    \end{align}
    where $Z={\wfomc(\Omega,w,n)}$ is the partition function. 
    \end{theorem}
    \begin{proof}[Sketch]
      The proof is a simple consequence of the fact that all the models agreeing
    with a count statistic $\bm n$ can be counted using cardinality constraints which
    agree with $\bm n$. Any such cardinality constraint correspond to a specific set
    of $(\kh)$ vectors. Hence, we can express arbitrary probability distributions over
    count statistics by picking real valued weights for $(\kh)$ vector. We defer the full proof to appendix.

  \end{proof}
    \begin{example}
      \label{coins}

      In the example proposed in \cite{complexMLNkuzelka2020}, they  model the distribution of a sequence of 4 coin tosses such that 
      the probability of getting an odd number of heads is zero and the probability of 
      getting an even number of heads is uniformly distributed.
      In order to model this distribution, we introduce a predicate $H(x)$ over a domain of $4$ elements, we also define $\Omega$ as $\top$. This means that every model of this theory is a model of $\Omega$. Notice that this distribution 
      cannot be expressed using symmetric weights, as symmetric weights can only express binomial 
      distribution for this language. But we can define weight function on $(\kh)$ vector. In this case $\bk=(k_0,k_1)$ such that $k_0+k_1=4$.  Since there are
      no binary predicates we can ignore $\bh$. Intuitively, $k_0$ is the
      number of tosses which are not heads  and $k_1$ is the number of tosses which are heads. If we define the weight function as $w(k_0,k_1) = 1 + (-1)^{k_1}$. Then by applying \eqref{eq:conting-distribution} we obtain the following
    probability distribution over the tosses: 
    \begin{align*}
    q(\Omega,(4,0))  &= \frac{\binom{4}{4}\cdot(1 + 1)}{16} = \frac18 \\ 
    q(\Omega,(3,1))  &= \frac{\binom{4}{3}\cdot(1 - 1)}{16}  = 0 \\ 
    q(\Omega,(2,2))  &= \frac{\binom{4}{2}\cdot(1 + 1)}{16} = \frac34 \\  
    q(\Omega,(1,3))  &= \frac{\binom{4}{1}\cdot(1 - 1)}{16} = 0 \\ 
    q(\Omega,(0,4))  &= \frac{\binom{4}{0}\cdot(1 + 1)}{16}  = \frac18  
    \end{align*}
which coincides with the distribution obtained by \cite{complexMLNkuzelka2020}. Notice, that such a distribution cannot be expressed through symmetric weight functions and obligates the use of a strictly more expressive class of weight functions. 

    \end{example}
    
    We are able to capture count distributions 
    without losing domain liftability.
    Furthermore, we do not introduce complex or even negative weights,
    making the relation between weight functions and probability rather intuitive. 
   
    \section{Conclusion}
    In this paper, we have presented a closed-form formula for FOMC
    of universally quantified formulas in FO$^2$ that 
    can be computed in polynomial time w.r.t. domain cardinality.
    From this, we are able to derive a closed-form expression for FOMC 
    in FO$^2$ formulas in Scott's Normal Form, extended with cardinality 
    constraints and counting quantifiers. These extended formulas are also
    computable in polynomial time, and therefore they constitute lifted
    inference algorithms for C$^2$. All the formulas are extended to cope
    with weighted model counting in a simple way, admitting a larger class
    of weight functions than symmetric weight functions. 
    All the results have been obtained using combinatorial principles, providing a
    uniform treatment to all these fragments.

    \section{Acknowledgements}
    We thank Alessandro Daniele
    and the anonymous reviewers for providing substantial help
    in improving the quality of this paper.

\bibliography{aaai22}  
\appendix
\section{APPENDIX}
\subsection{FOMC for Universal Formulas}

\ccflemma*
\begin{proof}[Proof of Lemma \ref{lemm:LI_PERM}]
  For any $ \bm{x'}\in X^{m}$, we have that 
  $\forall \bm{x} \Phi(\bm{x}) \rightarrow \forall
  \bm{x}\Phi(\bm{x'})$ is valid. Which implies that 
  $\forall \bm{x} \Phi(\bm{x}) \rightarrow \bigwedge_{\bx'\in X^m}\forall
  \bm{x}\Phi(\bm{x'})$ is also valid. Since $\forall$ and $\wedge$ commute,
  we have that $\forall \bx.\Phi(\bx) \rightarrow
  \forall\bx.\Phi(X)$. The viceversa is obvious since $\Phi(\bx)$ is
  one of the conjuncts in $\Phi(X)$. 
\end{proof}

% \begin{proposition} 
%   \label{prp:permutation-invariance-truth-assignment}
%   For every pure universal formula $\Phi(\bx)$, every permutation $\pi$
%   of $X$ and  every lifted interpretation $\tau$ for $\Phi(X)$, 
%   $\tau(\Phi(X)) = \tau_{\pi}(\Phi(X))$;
%   where $\tau_\pi(P(x_i,x_j,\dots)=
%   \tau(P(\pi(x_i),\pi(x_j),\dots)$, for every atom $P(x,y,\dots)$.
%   \end{proposition}

% \begin{proof}
%   If $\tau(\Phi(X))=0$ then $\tau(\Phi(\bx'))=0$ for some
%   $\bx'\in X^m$. This implies that $\tau_\pi(\Phi(\pi^{-1}(\bx')))=0$,
%   which implies that $\tau_\pi(\Phi(X))=0$. The proof
%   of the opposite direction follows form the fact that
%   $(\tau_\pi)_{\pi^{-1}}=\tau$. 
%   \end{proof}

In order to prove Theorem \ref{thm:mc-pure-univ-formulas}, we first introduce 
  the following notation and we also introduce
  Lemma \ref{lem:mc-Phi-C} and Lemma \ref{lem:join}.\\ 
  For any set of constants $C$ and any $2^u$-tuple $\bk=(k_0,\dots,k_{2^u-1})$
such that $\sum\bk=\card{C}$, let $\CC_\bk$ be any partition $\{C_i\}_{i=0}^{2^u-1}$
of $C$ such that $\card{C_i}=k_i$. 
We define  $\Phi(\mathbb{C}_{\bm{k}})$ as follows: 
\begin{equation}
  \label{eq:phi-c-k}
    \Phi(\CC_\bk) = \Phi(C) \wedge
\bigwedge_{i=0}^{2^u-1}\bigwedge_{c\in C_i}i(c)
\end{equation}

\begin{example}
Examples of $\CC_{(1,0,2,0)}$, on $C=\{a,b,c\}$ are 
$\{\{a\},\emptyset,\{b,c\},\emptyset\}$ and
$\{\{b\},\emptyset,\{a,c\},\emptyset\}$. 
\begin{align*}
    \Phi(\{ \{a\},\emptyset,\{b,c\},\emptyset\})  = \Phi(C) &\land \neg A(a) \land \neg R(a,a)\\
    &\land  A(b) \land \neg R(b,b)\\
    &\land A(c) \land \neg R(c,c)  \\
\end{align*}
Note there are ${3 \choose 1,0,2,0}=3$ such partitions, and all the $\Phi(\CC_\bk)$ for such partitions will have the same model count. These observations have been formalized in lemma \ref{lem:mc-Phi-C}
\end{example}

\begin{lemma}
  \label{lem:mc-Phi-C}
  Given a pure universal formula $\forall \bx\Phi(\bx)$ in FO$^2$, 
$    \mc(\Phi(C)) = \sum_{\bk}\binom{n}{\bk} \mc(\Phi(\mathbb{C}_{\bm{k}}))$, where $\mc(\alpha)$ denotes the 
model count of an arbitrary propositional formula $\alpha$.
\end{lemma}
\begin{proof}
  First let us show that the $\mc(\Phi(C))$ is independent from the
  specific choice of $\CC_\bk$ for every $\bk$.
  Let $\CC_\bk$ and $\CC'_\bk$, be two partitions with the same $\bk$.
  Notice that $\CC'_{\bk}$ can be obtained by applying some
  permutation on $C$ from $\CC_\bk$, hence 
  $$\mc(\Phi(\CC_\bk))=\mc(\Phi(\CC'_\bk))$$
  Furthermore notice that if $\CC_\bk$ is different from $\CC'_{\bk'}$
  then $\Phi(\CC_\bk)$ and $\Phi(\CC'_{\bk'})$ cannot be
  simultaneously satisfied. This implies that
  $$
  \mc(\Phi(C)) = \sum_{\bk}\sum_{\CC_\bk}\mc(\Phi(\CC_\bk))
  $$
  Since there are ${n \choose \bk}$ partitions of $C$, of the
  form $\CC_\bk$, then 
  $$
  \mc(\Phi(C)) = \sum_{\bk}{n\choose\bk}\mc(\Phi(\CC_\bk))
  $$
\end{proof}

\begin{lemma}
\label{lem:join}
For any partition $\CC_\bk=\{C_0,\dots,C_{2^u-1})$ 
$$\mc(\Phi(\CC_\bk)) = \prod_{\substack{c\neq d  \\ c,d \in C}} n_{i_c i_d}$$ 
where  for all $c,d\in C$, $0\leq i_c,i_d\leq 2^u-1$ are the indices such that 
$c\in C_{i_c}$ and $d\in C_{i_d}$. 
\end{lemma}

\begin{proof}
  $\Phi(\CC_\bk)$ can be rewritten in
  $$
  \bigwedge_{\{c,d\}\subseteq C \atop c\neq d}\Phi^{i_c,i_d}(\{c,d\})
  $$
  $\Phi^{i_c,i_d}(\{c,d\})$ is obtained by replacing each atom
  $P_j(c)$ with $\top$ if $bin(i_c)_j=1$ and $\bot$ otherwise
  and each atom $P_j(d)$ with $\top$ if
  $bin(i_d)_j=1$ and $\bot$ otherwise.
  Notice that all the atoms of 
  $\Phi^{i_c,i_d}(\{c,d\})$ contain both $c$ and $d$.
  Furthermore notice that if $\{c,d\}\neq\{e,f\}$ then
  $\Phi^{i_c,i_d}(\{c,d\})$ and $\Phi^{i_e,i_f}(\{e,f\})$ do not
  contain common atoms.
  Finally we have that $\mc(\Phi^{i_c,i_d}(\{c,d\}))=n_{i_ci_d}$.
  Hence
  $$
  \mc\left(\bigwedge_{c, d\in C\atop c\neq
      d}\Phi^{i_c,i_d}(\{c,d\})\right)= \prod_{\substack{c\neq d  \\
      c,d \in C}} n_{i_c i_d}
  $$ 
\end{proof}
Finally, we provide the following proof for Theorem \ref{thm:mc-pure-univ-formulas}.

\begin{proof} [Proof of Theorem \ref{thm:mc-pure-univ-formulas}]
    Notice that $\fomc(\Phi(\bx),n)=\mc(\Phi(C))$ for a set of constants
    $C$ with $\card{C}\ =n$. Therefore, by Lemma~\ref{lem:mc-Phi-C}, to prove
    the theorem it is enough to show that for all $\bk$,
    $\mc(\Phi(\CC_{\bk}))= \prod_{0 \leq i \leq j \leq 2^u-1}
    n_{ij}^{\bk(i,j)}$. By the Lemma~\ref{lem:join} we have that 
  $ \mc(\Phi(\CC_\bk)) = \prod_{c\neq d} n_{i_c i_d}$. Then:
  \begin{align*}
  \prod_{c\neq d} n_{i_c i_d} & = 
    \prod_{i} \prod_{\substack{c \neq d \\  c,d \in C_{i}}} n_{ii} \cdot \prod_{i< j}\prod_{\substack{c \in C_i \\ d \in C_j }} n_{ij}  \\
   & =  \prod_{i} n_{ii}^{k_{i} \choose 2} \cdot \prod_{i < j}  n_{ij}^{k_{i}k_{j}} =
   \prod_{0 \leq i \leq j < 2^u} n_{ij}^{\bk(i,j)}
  \end{align*}
  \end{proof}

    As a final remark for this section, notice that the computational cost of computing $n_{ij}$ is constant with respect to the domain cardinality. We assume the cost of multiplication to be constant. Hence, the computational complexity of computing  \eqref{eq:fomc-universal-form} depends on the domain only through the multinomial coefficients $\binom{n}{\bk}$ and the multiplications involved in $\prod_{ij} n^{\bk(i,j)}$. The computational cost of computing  $\binom{n}{\bk}$ is polynomial in $n$ and the total number of $\binom{n}{\bk}$  are $\binom{n+2^{u}-1}{2^{u}-1}$, which has
    $\left(\frac{e\cdot(n+2^{u}-1)}{2^{u}-1}\right)^{2^{u}-1}$ as an upper-bound. Also, the $\prod_{ij} n^{\bk(i,j)}$ term has  $O(n^2)$ multiplication operations. Hence, we can conclude that the \eqref{eq:fomc-universal-form} is computable in polynomial time with respect to the domain cardinality. 

%---------------------------------------------------%
\section{FOMC for Cardinality Constraints }
In the following, we provide some examples to better 
explain $(\kh)$ vectors.
\begin{example}
  To count the models of \eqref{eq:example} with
  the additional constraint that $A$ is balanced i.e.,
  \mbox{$\frac{n}{2}\leq\card{A}\leq\frac{n+1}{2}$}, we have to consider only the
  terms where $\bk$ is such \mbox{$\frac{n}{2}\leq
    \bk(A)\leq\frac{n+1}{2}$}.
  Equivalently, we should
  consider only the $\bk$ such that $\frac{n}{2}\leq k_2+k_3\leq
  \frac{n+1}{2}$. (Notice that $k_2$ is the number of elements that
  satisfy $A(x)$ and $\neg R(x,x)$ and $k_3$ is the number of elements
  that satisfy $A(x)$ and $R(x,x)$). 
  \end{example}

\begin{example}
  A graphical representation of the pair $\bk,\bh$ for the formula
  \eqref{eq:example} is provided in the following picture: 
  \begin{center}
    \begin{tikzpicture}
      \foreach \i in {0,...,3}{
        \node at (1.2*\i cm,0.75) {$k_\i$};
        \node at (-1,-\i cm) {$k_\i$};      
         \foreach \j in {\i,...,3}
          \node[rectangle,draw,minimum width=1.2cm, minimum height =
          1cm,inner sep = 1pt ]
          at (1.2*\j cm,-\i cm) (h\i\j)
          {$\begin{smallmatrix}h^{\i\j}_0 & h^{\i\j}_1 \\ h^{\i\j}_2 &  h^{\i\j}_3\end{smallmatrix}$};};
      % \node[rectangle,draw,minimum width=1.2cm, minimum height =
      %     1cm,inner sep = 1pt,fill=red!10] at (h12)
      %     {$\begin{smallmatrix}h^{12}_0 & h^{12}_1 \\ h^{12}_2 &  h^{12}_3\end{smallmatrix}$};
      \end{tikzpicture}
    \end{center}
  This configuration represent the models in which a set $C$ of $n$ 
  constants are partitioned in four sets $C_0,\dots,C_3$, each $C_i$ containing
  $k_i$ elements (hence $\sum k_i=n$). Furthermore, for each pair $C_i$ and $C_j$ the relation
  $D^{ij}=C_i\times C_j$ is  partitioned in 4 sub relations
  $D^{ij}_0,\dots,D^{ij}_3$ where each $D^{ij}_v$ contains $h^{ij}_v$
  pairs (hence $\sum_vh_v^{ij}=\bk(i,j)$). For instance if the pair $(c,d)\in D^{12}_2$ it means that
  we are considering assignments that satisfy
  $\neg A(c) \land R(c,c) \land A(d) \land \neg R(d,d) \land R(c,d)
  \land \neg R(d,c)$.
  \end{example}

\section{Weighted First Order Model Counting }
All the FOMC formulas introduced in the paper can be 
extended to WFOMC by defining a real-valued function on $(\bk,\bh)$ and adding it 
as a multiplicative factor, for instance if $\Phi$ is in SNF, as given in equation \eqref{eq:simple-scott}, then it's WFOMC can be defined as follows:
\begin{align*}
  \mathrm{WFOMC}(\Phi,n) = \sum_{\bk,\bh}(-1)^{\sum_{i}\bk(P_i)}w(\bk,\bh)F(\bk,\bh,\Phi')
\end{align*}
where $w(\bk,\bh)$ is a real-valued weight function and $\Phi'$ is the transformed formula as described in Theorem \ref{thm:fomc-scott-form}.
\subsection{Symmetric Weight Functions}

\begin{theorem}
For all $\Phi$ in C$^2$ and for arbitrary cardinality 
constraint
$\rho$, symmetric-WFOMC
can be obtained from FOMC by defining the following weight function:
\begin{align*}
  w(\bk,\bh) & = \prod_{P\in\L}
  w(P)^{(\bk,\bh)(P)}\cdot 
  \bar{w}(P)^{(\bk,\bh)(\neg P)}
\end{align*}
where $w(P)$ and $\bar{w}(P)$ are real valued weights on predicate $P$ and it's negation respectively.
\end{theorem}

\begin{proof}
The proof is a consequence of the observation that
$F(\bk,\bh,\Phi)$ is the number of models of $\Phi$ that
contain $\bk(P)$ elements that satisfy $P$ if $P$ is unary,
and $(\bk,\bh)(P)$ pairs of elements that satisfy $P$, if $P$ is
binary. 
\end{proof}

\subsection{Expressing Count Distributions}
In the following we provide the proof for Theorem \ref{thm:Counting_WFOMC}.

  \begin{proof}[Proof of Theorem \ref{thm:Counting_WFOMC}]
    Since $\Omega$ is a FO$^2$ formula, then we can compute FOMC as
    follows: 
    $$\fomc(\Omega,n) =
    \sum_{\bk,\bh}F(\bk,\bh,\Omega)
    $$
    Let us define $w(\bk,\bh)$ for each $\bk,\bh$ as follows: 
    $$
    w(\bk,\bh)=
    \frac1{F(\bk,\bh,\Omega)}\sum_{\substack{
      \omega\models\Omega \\
      N(\alpha_1,\omega)_1 = (\bk,\bh)(P_1) \\ \dots \\
      N(\alpha_m,\omega)_m = (\bk,\bh)(P_m)}}
  p_{\Phi,\Omega}(\omega)
  $$
  This definition implies that the partition function $Z$ is equal to
  1. Indeed:
  \begin{align*}
    Z & = \wfomc(\Omega,w,n) \\
      & = \sum_{\bk,\bh}w(\bk,\bh)\cdot F(\bk,\bh,\Omega) \\
      & = \sum_{\bk,\bh}
          \sum_{\substack{
      \omega\models\Omega \\
      N(\alpha_1,\omega)_1 = (\bk,\bh)(P_1) \\ \dots \\
      N(\alpha_m,\omega)_m = (\bk,\bh)(P_m)}}
    p_{\Phi,\Omega}(\omega) \\
      & = \sum_{\omega\models\Omega}
          \sum_{\substack{
      \bk,\bh \\
      N(\alpha_1,\omega)_1 = (\bk,\bh)(P_1) \\ \dots \\
      N(\alpha_m,\omega)_m = (\bk,\bh)(P_m)}}
    p_{\Phi,\Omega}(\omega) \\
      & = \sum_{\omega\models\Omega}p_{\Phi,\Omega}(\omega) \\ 
  & = 1   
  \end{align*}
  Hence,
  \begin{align*}
  q_{\Phi}(\Omega,\bm n) & =
  \sum_{(\bk,\bh)(P_i)=n_i}
                           F(\bk,\bh,\Omega)\cdot w(\bk,\bh) \\
                         & = \sum_{(\bk,\bh)(P_i)=n_i}
    \sum_{\substack{
      \omega\models\Omega \\
      N(\alpha_1,\omega)_1 = (\bk,\bh)(P_1) \\ \dots \\
      N(\alpha_m,\omega)_m = (\bk,\bh)(P_m)}}
    p_{\Phi,\Omega}(\omega) \\
                         & = 
    \sum_{\substack{
      \omega\models\Omega \\
      N(\alpha_1,\omega)_1 = n_1 \\ \dots \\
      N(\alpha_m,\omega)_m = n_m}}
    p_{\Phi,\Omega}(\omega) \\
  \end{align*}
  
  \end{proof}

%%% Local Variables:
%%% mode: latex
%%% TeX-master: "lwmc_AAAI"
%%% End:

\end{document}